\documentclass[twocolumn,showpacs,preprintnumbers,prl,fleqn]{revtex4}
\usepackage{graphicx}
\usepackage{dcolumn}
\usepackage{bm}

\begin{document}

\title{Collective character of spin excitations in a system of Mn$^{2+}$ spins coupled to a
two-dimensional electron gas}
\author{F.J. Teran, M. Potemski, D.K. Maude, D. Plantier, A.K. Hassan, and A. Sachrajda}
\altaffiliation[Permanent address ]{Institute for Microstructural
Sciences, National Research Council, Ottawa, Canada K1A 0R6}
\affiliation{Grenoble High Magnetic Field Laboratory, Max Planck Institut f\"{u}r Festk%
\"{o}rperforschung\\
and Centre National de la Recherche Scientifique, BP 166, 38042
Grenoble Cedex 9, France.}
\author{Z. Wilamowski, J.\ Jaroszynski, T. Wojtowicz, and G. Karczewski}
\affiliation{Institute of Physics, Polish Academy of Sciences,
02668 Warsaw, Poland}
\date{\today }

\begin{abstract}
We have studied the low energy spin excitations in n-type CdMnTe
based dilute magnetic semiconductor quantum wells. For magnetic
fields for which the energies for the excitation of free carriers
and Mn spins are almost identical an anomalously large Knight
shift is observed. Our findings suggests the existence of a
magnetic field induced ferromagnetic order in these structures,
which is in agreement with recent theoretical predictions [J.
K{\"o}nig and A. H. MacDonald, submitted Phys. Rev. Lett. (2002)].
\end{abstract}

\maketitle


Although the existence of a ferromagnetic phase in diluted
magnetic semiconductors (DMS) is experimentally well
established\cite{Story,Haury,Ohno92,Ohno96}, the physical origin
of this phenomenon is far from being well
understood\cite{Dietl,Bhatt,Konig00}. The RKKY approach
\cite{Kittel}, which successfully explains the ferromagnetism
observed in magnetic metals, cannot easily be applied to the case
of magnetic semiconductors which are typically composed of a
dilute subsystem of localized magnetic spins and an even more
dilute gas of free carriers. On the other hand, the early Zener
model\cite{Zener} of ferromagnetism driven by the exchange
interaction between free carriers and localized magnetic moments
provides a rough estimate of the observed critical ferromagnetic
temperatures in DMS materials\cite{Dietl}. The Zener model
nevertheless neglects important effects related to the character
of the ferromagnetic order in these systems, possibly mediated by
the itinerant nature of the free carriers spins\cite{Konig00}.
This suggests that diluted magnetic semiconductors show a new
class of ferromagnetism, which however, remains to be
experimentally verified.

In this letter, we report on the investigations of spin
excitations in a model DMS structure, namely very diluted
Mn$^{2+}$ ions coupled to an electron gas both of which are
confined in a CdMnTe quantum well structure. Our key experimental
results rely on an accurate and local probing of Mn$^{2+}$ spin
excitations using the electron paramagnetic resonance (EPR)
technique. At low magnetic fields ($B<4 T$) and sufficiently high
temperatures ($T\sim 4.2 K$), the investigated structure show all
the attributes of a system composed of two paramagnetic
subsystems: localized Mn$^{2+}$ ions and an electron gas, coupled
via the $s-d$\ exchange interaction. The main experimental finding
reported here is the change in the line-shape and the giant shift
of the Mn$^{2+}$ resonance observed under specific conditions when
the spin polarization of free carriers is induced by the
application of the high magnetic field and when at the same time
the energies of the mean field spin excitations of electron's and
Mn$^{2+}$ spins are comparable. This finding can be considered as
a positive test for the recent theory of ferromagnetism in DMS
materials\cite{Konig00}. The observed changes in the spin
resonance spectrum indicate the formation of two macroscopic
moments characteristic for each spin subsystem, which are
efficiently coupled via their transverse components. Although the
expected zero-field ferromagnetic critical temperature for the
investigated n-type structures is expected to be very low
($\lesssim 5mK$) \cite{Konig02}, our data suggest the appearance
of ferromagnetic order in these systems when the spin polarization
is forced by the application of a magnetic field.

The two samples A and B used for investigations were 10nm-thick
CdMnTe/CdMgTe single quantum well structures with a modulation
doping on one side of the quantum well (QW). Both samples have
been characterized by conventional magneto-luminescence
measurements whereas sample B has also been intensively studied
with magneto-transport and cyclotron resonance absorption
measurements \cite{Teran}. In sample A, the estimated electron
sheet density is n$_{e}\simeq $1$\times $10$^{11}$cm$^{-2}$
(corresponding 3D concentration n$_{e}^{3D}\simeq $1$\times $10$^{17}$cm$%
^{-3}$), and an effective Mn$%
^{2+}$ concentration in the QW x$_{eff}\simeq $0.2{\%}\ (n$_{Mn}^{3D}\simeq $%
3$\times $10$^{19}$cm$^{-3}$). The parameters of sample B are more
precisely determined: n$_{e}$=5.95$\times $10$^{11}$%
cm$^{-2}$, (n$_{e}^{3D}\simeq $6$\times $10$^{17}$cm$^{-3}$), mobility $%
\mu $=60000 cm$^{2}$/Vs and x$_{eff}$=0.3%
{\%} (n$_{Mn}^{3D}\simeq $4.4$\times $10$^{19}$cm$^{-3}$).

The spin excitations have been probed using Raman scattering and
resistively detected multi-frequency EPR. Raman scattering allows
to probe the spin flip transitions of both band electrons and
Mn$^{2+}$ ions but can only be easily applied for samples with low
electron concentrations, which show sharp (exciton like)
absorption lines and therefore a large resonant enhancement of the
scattering signal. Traditional EPR techniques give a higher
resolution, but are difficult to apply in our case due to the
small number of spins. Here we locally probe the Mn$^{2+}$ EPR in
CdMnTe quantum wells via detecting the microwave induced changes
in the longitudinal resistance using microwave sources (operating
at 95 GHz and 230 GHz (Gunn diodes) and in the range of 64-95 GHz
(carcinotrons)).

In a first approach, we consider the system to be composed of two
paramagnetic subsystems: two-dimensional conduction electrons with
extended wave functions and localized $3d^{5}$ states of Mn$^{2+}$
ions, which interact via the $s-d$\ exchange interaction.
Reasoning in terms of the conventional mean field approximation,
we expect the energies of spin excitations for electrons,
$E_{e}^{S}$, and Mn$^{2+}$ ions, $E_{Mn}^{S}$\ to be:
$E_{e}^{S}=E_{e}^{Z}+\Delta _{E}$ and
$E_{Mn}^{S}=E_{Mn}^{Z}+K_{E}$,
where, correspondingly for electrons and Mn$^{2+}$ ions,  $E_{e}^{Z}$ and $%
E_{Mn}^{Z}$ are the energies in the absence of the $s-d$ exchange
interaction, whereas $\Delta _{E}$ and $K_{E}$ denote the mean
field exchange terms. We consider only the relevant spin
excitations with $|\Delta m_{S}|=1$, (m$_{S}$ is the quantum
number associated with the projection of the spin along the
external magnetic field direction) and in a first approximation
assume that $E_{e}^{Z}$ and $E_{Mn}^{Z}$ are
given by the usual Zeeman terms: $E_{e}^{Z}=g_{e}\mu _{B}B$ $\ $and $%
E_{Mn}^{Z}=g_{Mn}\mu _{B}B$ (throughout this paper we take $g_{e}$%
=-1.64 and $g_{Mn}$=2.007 for the electronic and Mn$^{2+}$
g-factors, respectively \cite{Sirenko,Deigen}). Including the mean
field $s-d$\ exchange, the characteristic spin excitations of our
system can be expressed as follows:

\begin{equation}
\label{eq1}
E_{e}^{S}=E_{e}^{Z}+\Delta _{E}=g_{e}\mu _{B}B+J_{sd} n_{Mn}^{3D}\frac{5}{2}%
\sigma _{Mn}^{z}\qquad
\end{equation}

\begin{equation}
\label{eq2}
E_{Mn}^{S}=E_{Mn}^{Z}+K_{E}=g_{Mn}\mu _{B}B+J_{sd} n_{e}^{3D}\frac{1}{2}%
\sigma _{e}^{z}
\end{equation}

\noindent Here, $\frac{1}{2}\sigma _{e}^{z}$ and
$\frac{5}{2}\sigma _{Mn}^{z}$ denote mean values of the S$_{z}$
spin components, correspondingly, for electrons (with spin S=1/2)
and Mn$^{2+}$ ions (with spin S=5/2), and $J_{sd} $ is the
exchange constant. $\sigma _{Mn}^{z}$ and $\sigma _{e}^{z}$ should
be identified with the normalized (to unity) spin polarization of
Mn$^{2+}$ ions and electrons, respectively. Rewriting the above
equations in the following form:
\begin{equation}
\label{eq3}
E_{e}^{S}=g_{e}\mu _{B}(B+\frac{\Delta _{E}}{g_{e}\mu
_{B}})=g_{e}\mu _{B}(B+\Delta _{B})
\end{equation}

\begin{equation}
\label{eq4} E_{Mn}^{S}=g_{Mn}\mu _{B}(B+\frac{K_{E}}{g_{Mn}\mu
_{B}})=g_{Mn}\mu _{B}(B+K_{B})
\end{equation}
there is an evident analogy between physics of DMS materials and
that of nuclear spins in metals coupled to the electrons spins via
the hyperfine interaction. Whereas the Overhauser shift, $\Delta
_{B}$ (energy, $\Delta _{E}$)  is expected to be significant in
our structures ($\Delta _{E}^{\max }=J_{sd}
n_{Mn}^{3D}\frac{5}{2}\sim 1.5 meV$), the Knight shift $K_{B}$
(energy, $K_{E}$)is much more subtle: $K_{E}^{\max }/\Delta
_{E}^{\max
}=n_{e}^{3D}/5n_{Mn}^{3D}$, so that $K_{E}^{\max }=J_{sd} n_{e}^{3D}\frac{1}{2}%
\sim 1-4\mu eV$.

\begin{figure}
\includegraphics[width=0.7\linewidth,angle=0,clip]{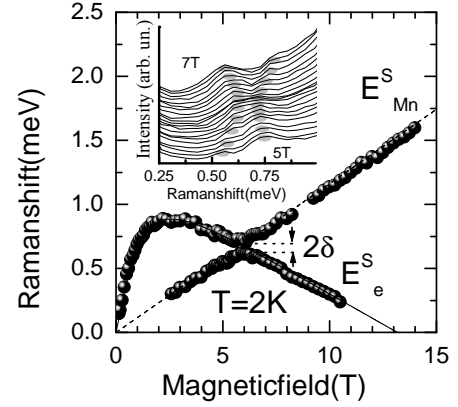}
\caption{\label{fig1} Raman shift for the spin-flip transitions
for electrons and Mn$^{2+}$ ions for sample A. The expected
behavior for non interacting subsystems calculated as described in
the text is indicated by the solid and dashed lines. The inset
shows the Raman spectra (every 0.1T) in the region of the avoided
crossing. The shaded regions are a guide to the eye.}
\end{figure}

An overview of the relevant spin excitations for the investigated
system can be obtained from Raman scattering spectra. The measured
energies of the excitations, which can be easily identified with
the spin-flip transitions for electrons and Mn$^{2+}$ ions are
shown in Fig.\ref{fig1}. The solid line represents the $E_{e}^{S}$
versus $B$ dependence according to Eq(\ref{eq1}) assuming that the
spin polarization of Mn$^{2+}$ is given by the modified Brillouin
function: $\sigma _{Mn}^{z}=B_{5/2}(g_{Mn}\mu _{B}B/k(T+T_{0}))$ ,
with two adjustable parameters: the saturated value of the
exchange term $\Delta _{E}^{\max }=J_{sd}
n_{Mn}^{3D}\frac{5}{2}=1.25 meV$ and $T_{0}=0.12 K$ which
phenomenologically accounts for the small antiferromagnetic correction for the Mn$%
^{2+}$ ensemble. Note, that the signs of the electron g-factor and
the exchange constant in CdMnTe are such that the effective
electron spin splitting results from the competition between the
intrinsic Zeeman term and
the $s-d$\ exchange contribution (we use the convention that $E_{e}^{Z}<0$ and $%
\Delta _{E}>0$). The dashed, line in Fig.\ref{fig1} is the
predicted linear variation of  the Mn$^{2+}$ spin excitations when
neglecting the small Knight shift and possible subtle corrections
to the Mn$^{2+}$ spin Hamiltonian related to hyperfine interaction
and/or crystal field effects which are in any case beyond the
resolution of the Raman scattering data. This simple model
satisfactorily reproduces the observed spin excitation energies
except in the region around $\sim 5 T$ where $E_{e}^{S}\sim
E_{Mn}^{S}$ in sample A. A clear indication of the avoided
crossing of electron and Mn$^{2+}$ spin excitation with the
characteristic repulsion energy $2\delta \approx 0.08 meV$ is
observed (see inset to Fig.\ref{fig1}). This is our key
experimental observation, which will be further elucidated using
the more precise EPR spectroscopic tool.

\begin{figure}
\includegraphics[width=0.7\linewidth,angle=0,clip]{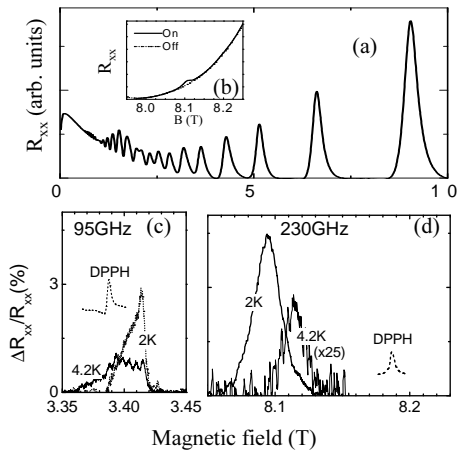}
\caption{\label{fig2} (a)Longitudinal resistance R$_{xx}$ as a
function of magnetic field. (b) R$_{xx}$ with and without 230 GHz
microwave illumination. (c) and (d)Typical resistively detected
EPR spectra. The EPR spectra of the DPPH g-factor marker is shown
for comparison.}
\end{figure}

Sample B, chosen for the resistively detected EPR measurements,
has been prepared in the form of a 0.5 x 1 mm Hall bar. As
illustrated in Fig.\ref{fig2}(a), it shows typical
magneto-transport properties for a 2-DEG. Using the available
microwave sources, the resonant change in the longitudinal
resistance induced by microwave illumination and identified with
the Mn$^{2+}$ EPR signal, has been measured at magnetic fields B
$\sim $ 2.4, 2.7, 3.5 and 8.1 T (correspondingly, in the vicinity
of filling factors
$\nu $=10,9,7 and 3). The measurements around B$\sim $%
8.1T correspond to the case when the energy of the spin
excitations of the $E_{e}^{S}$, and $E_{Mn}^{S}$ subsystems are
expected to be almost identical in sample B. An example of the
measured resonant increase of the R$_{xx}$ around 8.1 T ($\nu \sim
3$) is illustrated in Fig.\ref{fig2}(b), where an expanded view of
$(R_{xx})$ is shown with and without microwave illumination at 230
GHz. In order to calibrate exactly the magnetic field value, a
small amount of a g-factor marker diphenyl-picryl-hydrazyl (DPPH)
with g=2.0036 is placed close to the sample and its EPR spectrum
is simultaneously measured using carbon bolometer mounted below
the sample. Typical EPR spectra obtained by subtracting the
resistance measured with and without microwave illumination ($\Delta $R$%
_{xx})$ when using 95 GHz and 230 GHz microwave sources are shown
in Fig.\ref{fig2}(c) and (d) for two different temperatures.

The spectrum measured at 95 GHz and 4.2K can be recognized as the
typical signal of paramagnetic Mn$^{2+}$ ions. It shows six
relatively well pronounced components which result from the
hyperfine interaction between the spin of the Mn 3d$^{5}$
electrons and the Mn nuclear spin (I=5/2). The estimated hyperfine
splitting (constant) is 52 G ($A_{Mn}=49\times 10^{-4}cm^{-1}$).
The smaller signal on the low field side of the main spectrum
(which characteristically vanishes at lower temperatures) is due
to the fine splitting which results from crystal field effects
related to the strain present in the quantum well. The estimated
value of the fine constant is $D=47\times 10^{-4} cm^{-1}$. The
amplitude and shape of the resonance measured at low magnetic
fields depends strongly on temperature in the range between 4.2
and 2 K (see Fig.\ref{fig2}(c)). At T=2 K, as expected only the
main fine component is observed but surprisingly enough the
amplitudes of the hyperfine satellites are significantly modified
and/or the spectrum shape is significantly changed. This unusual
spectral shape at T = 2K either results from the effect of
dynamical nuclear polarization or indicates that the measured
resonance can no longer be attributed to paramagnetic Mn$ ^{2+}$
centers which then can be considered a first signature of a change
of the magnetic phase of the Mn$^{2+}$ ensemble at low
temperatures even at low magnetic fields.

\begin{figure}
\includegraphics[width=0.7\linewidth,angle=0,clip]{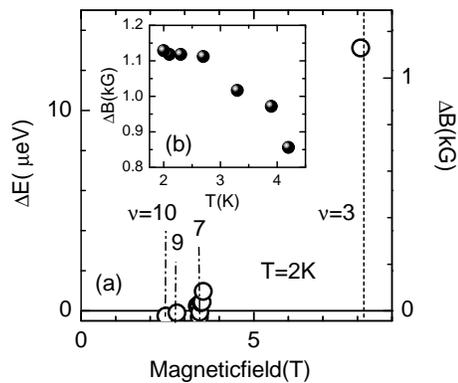}
\caption{\label{fig3} (a) Measured shift of the Mn$^{2+}$ EPR line
from the expected behavior. (b) Temperature dependence of the
large shift observed close to filling factor $\nu =3$.}
\end{figure}

The EPR spectra measured at 230 GHz around B=8T are distinctly
different (Fig.\ref{fig2}(d)). No trace of hyperfine splitting is
found. Instead the spectrum represents a symmetric, relatively
narrow single line. What is even more significant is that the
position of this line is shifted far from the value expected from
a simple linear extrapolation. This can clearly be seen in the raw
data from the respective positions of Mn$^{2+}$ and DPPH
resonances measured at 95 and 230 GHz shown in Figs.\ref{fig2} (c)
and (d), and is more clearly illustrated in Fig.\ref{fig3}, where
the difference between the measured resonance energies (magnetic
field position) and those given by $E_{Mn}^{S}=g_{Mn}\mu _{B}B-4D$
as expected for the low temperature paramagnetic Mn$^{2+}$
resonance are plotted for the data obtained at 2K at different
magnetic fields. While the above formula correctly reproduces the
resonance position at low fields, a very large shift of the
resonance ($\sim 1100 G$) towards lower fields can be deduced for
the 230 GHz spectra. As shown in Fig.\ref{fig2} (d) and
Fig.\ref{fig3}, the 230 GHz-resonance position is remarkably
sensitive to temperature. A shift of $\sim 300 G$ is observed when
the temperature is increased from 2 K to 4.2 K. At first sight one
might think that the observed change in the resonance position can
be simply related to a conventional Knight shift. However, the
saturated value of the Overhauser shift in this sample has been
precisely determined from the low field transport measurements
($\Delta _{E}^{\max }=1.65 meV$)\cite{Teran}. Thus the expected
amplitudes of Knight shift are $K_{E}=\Delta _{E}^{\max
}(n_{e}^{3D}/5n_{Mn}^{3D})\sigma _{e}^{z}=4.5\sigma _{e}^{z}$
$[\mu eV]$, where the maximum electron spin polarization can be
estimated from Landau level filling factors $\sigma _{e}^{z}\leq
1/\nu $ (for odd filling factors). Therefore, around B=8 T (where
$\nu \sim 3): K_{E}\leq 1.5\mu eV$ (or $K_{B}\leq 130 G$) and for
the case of experiments at low magnetic fields (where $\nu \geq
7$): $K_{E}\leq 0.7\mu eV$ (or $K_{B}\leq 40 G$). As can be seen
in Fig.\ref{fig3}, in the range of low magnetic fields the
measured resonance positions are consistent with the expected
Knight shift but this conventional effect is clearly unable to
account for either the observed position of the 230GHz resonance
or for its temperature dependence.

\begin{figure}
\includegraphics[width=0.7\linewidth,angle=0,clip]{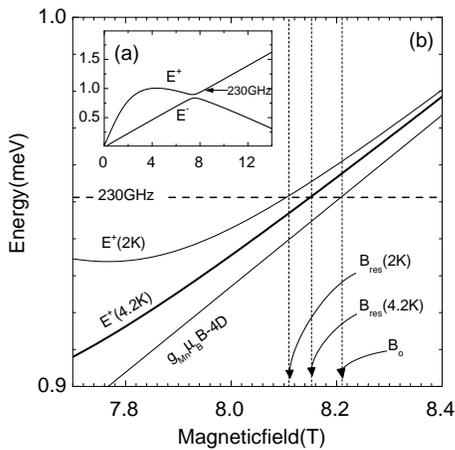}
\caption{\label{fig4} (a) Predicted spin excitations using the
phenomenological perturbation model as described in the text. (b)
Expanded view of the avoided crossing region.}
\end{figure}

Phenomenologically the position of the 230 GHz resonance can be
explained by taking into account the effect of the interaction
between the mean field modified $E_{e}^{S}$ and $E_{Mn}^{S}$ spin
excitations of the 2D electron and the Mn$^{2+}$ ensembles,
respectively. This manifest itself as the avoided crossing of the
corresponding excitation energies as already indicated by our
Raman scattering data for sample A. To simulate such an effect, we
introduce the coupling between $E_{e}^{S}$ and $E_{Mn}^{S}$
excitations using a simple perturbation approach and obtain:
$E^{\pm}=\frac{1}{2}(E_{e}^{S}+E_{Mn}^{S})\pm \frac{1}{2}
\sqrt{(E_{e}^{S}-E_{Mn}^{S})^{2}+4\delta ^{2}}$ for the new
coupled modes ($2\delta$ is the characteristic repulsion energy).
The expected $E^{\pm}$ energies for sample B are shown in the
Fig.\ref{fig4}(a), using a value of $\delta $=0.03 meV for the
adjustable interaction parameter. To derive the unperturbed modes
we have neglected the small Knight shift corrections assuming
$E_{Mn}^{S}=g_{Mn}\mu _{B}B-4D$ and $E_{e}^{S}$ given by
Eq.(\ref{eq1}) with the Mn$^{2+}$ polarization described by a
Brillouin function and the relevant parameters obtained for this
sample from low field transport measurements ($\Delta _{E}^{\max
}=1.65 meV $, $T_{0}=0.18 K$). Reasoning in terms of the mean
field approach we would expect the 230GHz resonance to occur at a
magnetic field just after the supposed crossing point of the
$E_{e}^{S}$ and $E_{Mn}^{S}$ energies. However, the effective
resonating branch at 230GHz turns out to be the E$^{+}$ coupled
mode. As emphasized in Fig.\ref{fig4}, our simple model explains
the actual resonance position and also accounts for the
temperature
dependence (with increasing temperature the crossing point for $E_{e}^{S}$ and $%
E_{Mn}^{S}$ excitations shifts towards lower fields). The
correspondence with the experimental data good. The predicted
resonance position is about 1 kG from the extrapolated value from
the low field data and the temperature driven shift is of about
400 G in the range between 2 K to 4.2 K.

To further pursue the phenomenological interpretation of our
experimental data, we refer to the theory of ferromagnetic
resonance and consider the possibility that, when a spin
polarization of Mn$^{2+}$and 2D electrons subsystems is induced by
magnetic fields, each spin subsystem may constitute a collective
(macroscopic) magnetic moment. Such collective modes are expected
to efficiently interact via the transverse components, leading to
mode repulsion with a characteristic interaction energy $(\Delta
_{E}K_{E})^{1/2}$ \cite{Konig02}. For sample B, we have determined
$\Delta_{E}=1.65 meV$ from low field transport \cite{Teran} and
deduce that $K_{E}\lesssim 1.5\mu eV$, the equality occurring for
the case where we have an ideal spin polarization of the 2DEG at
$\nu =3$. Therefore, in the upper limit $(\Delta
_{E}K_{E})^{1/2}=0.05 meV$ which is in fair agreement with the
experimentally found interaction parameter $\delta =0.03meV$. This
agreement is even better for sample A with lower electron
concentration, for which the 2DEG is very likely fully polarized
at the (anti)crossing point ($B\simeq6T$, $\nu =0.7$). There, we
expect $(\Delta _{E}K_{E})^{1/2}=0.03 meV$ and measure
$\delta\simeq0.04meV$(see Fig.\ref{fig1}).

Summarizing, we have investigated a very diluted magnetic
semiconductor system of Mn$^{2+}$ ions coupled to a 2DEG in CdMnTe
quantum well structures. The observed avoided crossing of the
Mn$^{2+}$ and the electron spin excitations is in agreement with
recent theoretical predictions \cite{Konig02} and is a signature
of the collective character of the spin excitations in DMS with
free carriers.

\begin{acknowledgments}
We thank J. K{\"o}nig and A.H. MacDonald for showing us their
theoretical results prior to publication. Support from INTAS
99-01146, Polonium, EU-SPINOSA-IST-2001-33334., and
PBZ-KBN-044/P03/2001 grants is acknowledged.

\end{acknowledgments}

\bigskip

\bibliography{TeranEPR}

\end{document}